\documentclass{rnaastex}
\usepackage[utf8]{inputenc}
\usepackage{bm}
\usepackage{amsmath}
\usepackage{booktabs}
\usepackage{graphicx}
\usepackage{hyperref}

\newcommand{\Msun}{\ifmmode {M_{\odot}}\else${M_{\odot}}$\fi}
\newcommand{\amin}{\ifmmode {^{\prime}}\else$^{\prime}$\fi}
\newcommand{\degree}{\ifmmode {^\circ}\else$^\circ$\fi}

\shorttitle{The Serendipitous Pulsar J0802$-$0955}
\shortauthors{Andrews et al.}

\begin{document}

\title{A Serendipitous Pulsar Discovery in a Search for a Companion to a Low-Mass White Dwarf}

\correspondingauthor{Jeff J.\ Andrews}
\email{andrews@physics.uoc.gr}

\author[0000-0001-5261-3923]{Jeff J.\ Andrews}
\affiliation{Foundation for Research and Technology-Hellas, 
100 Nikolaou Plastira St., 
71110 Heraklion, Crete, Greece}
\affiliation{Physics Department \& Institute of Theoretical \& Computational Physics, 
P.O Box 2208, 
71003 Heraklion, Crete, Greece}

\author[0000-0001-7077-3664]{Marcel A.~Ag\"ueros}
\affiliation{Department of Astronomy, Columbia University, 550 West 120th Street, New York, NY 10027, USA}

\author[0000-0002-1873-3718]{Fernando Camilo}
\affiliation{SKA South Africa, Pinelands 7405, South Africa}

\author[0000-0001-6098-2235]{Mukremin Kilic}
\affiliation{Department of Physics and Astronomy, University of Oklahoma, 440 W.\ Brooks St., Norman, OK, 73019, USA}
\affiliation{Institute for Astronomy, University of Edinburgh, Royal Observatory, Blackford Hill, Edinburgh EH9 3HJ, UK}

\author[0000-0002-8655-4308]{Alex Gianninas}
\affiliation{Department of Physics and Astronomy, University of Oklahoma, 440 W.\ Brooks St., Norman, OK, 73019, USA}
\affiliation{American Public University System, 111 W.\ Congress St., Charles Town, WV 25414, USA}

\author[0000-0002-4462-2341]{Warren Brown}
\affiliation{Smithsonian Astrophysical Observatory, 60 Garden St., Cambridge, MA 02138, USA}

\author[0000-0003-3944-6109]{Craig Heinke}
\affiliation{Department of Physics, University of Alberta, CCIS 4-181, Edmonton, AB T6G 2E1, Canada}

\section{} 

We report the discovery of a previously unidentified pulsar as part of a radio campaign to identify neutron star companions to low-mass white dwarfs (LMWDs) using the Robert C.\ Byrd Green Bank Telescope (GBT). PSR J0802$-$0955, which is coincident with the position of a WD with a mass of 0.2 \Msun\ \citep{gianninas15}, has a pulse period of 571 ms. Because of its relatively long pulse period, the lack of radial velocity (RV) variations in the radio data, and GBT's large beam size at the observing frequency of 340 MHz, we conclude that PSR J0802$-$0955 is unassociated with the LMWD at roughly the same position and distance.  

LMWDs are typically found in binary systems \citep{marsh95}, as the Universe is not old enough for lower mass stars to evolve completely and form isolated WDs with masses $\lesssim$0.45 \Msun\ \citep[although see][]{kilic07}. The Extremely Low Mass WD survey 
\citep[e.g.,][]{brown16} seeks to identify a sample of LMWDs in binary systems. Of the 88 LMWDs observed, at least 85\% show RV variations indicating they have unseen binary companions. While most LMWD companions are WDs, some may be neutron stars \citep{andrews14, boffin15, brown16}. Yet, none of the previous searches for neutron star companions to LMWDs using both radio and X-ray telescopes have identified such a system \citep{van_leeuwen07, agueros09a, agueros09b}.

To search for pulsar companions to LMWDs identified in the Extremely Low Mass WD survey, we engaged in a radio wave follow-up campaign using the GBT (Andrews et al., in prep). Data were folded and searched using {\tt PRESTO} \citep{ransom02, ransom03}\footnote{\href{https://www.cv.nrao.edu/~sransom/presto/}{https://www.cv.nrao.edu/~sransom/presto/}}. As part of this campaign (GBT/12A$-$431, GBT/14A$-$438, GBT/14B$-$347), we identified a pulsar with a spin period of 571.2564$\pm$0.0004 ms and a dispersion measure of 21.3 pc cm$^{-3}$ with 60-$\sigma$ confidence in a position coincident with the LMWD SDSS J080250.13$-$095549.8. A check of the Australia Telescope National Facility Pulsar Catalog \citep{manchester05} shows no known pulsar at this location. Our initial 75 minute discovery observation of this pulsar (30 May, 2014; 340 MHz central frequency, 100 MHz bandwidth; 4096 channels, GUPPI backend) was not long enough to span an appreciable fraction of the 13.1-hour orbit, as implied by the optical RV curve, to determine whether this pulsar is associated with the LMWD. 

\section{}

We obtained a 6-hour follow-up observation with the GBT (GBT/16A$-$324; 28 April, 2016; 340 MHz central frequency, 100 MHz bandwidth; 4096 channels; GUPPI backend) to search for Doppler shifts in the pulse period indicative of binarity. Using {\tt PRESTO}, we derive pulse periods for each of 10 separate $\approx$35-minute subdivisions of our total 6 hour observation. PSR J0802$-$0955 shows a stable pulse period with no Doppler variations down to a precision of $\sim$10$^{-6}$ s for individual 35-minute subdivisions. Based on the optical RV curve from the LMWD, a putative 1.4 \Msun\ NS companion would show modulations to the spin period of $\sim$10$^{-4}$ s over the length of our observation. We therefore conclude that PSR J0802$-$0955 is most likely an isolated, field pulsar. 

\begin{figure}
\begin{center}
\includegraphics[width=\columnwidth,angle=0]{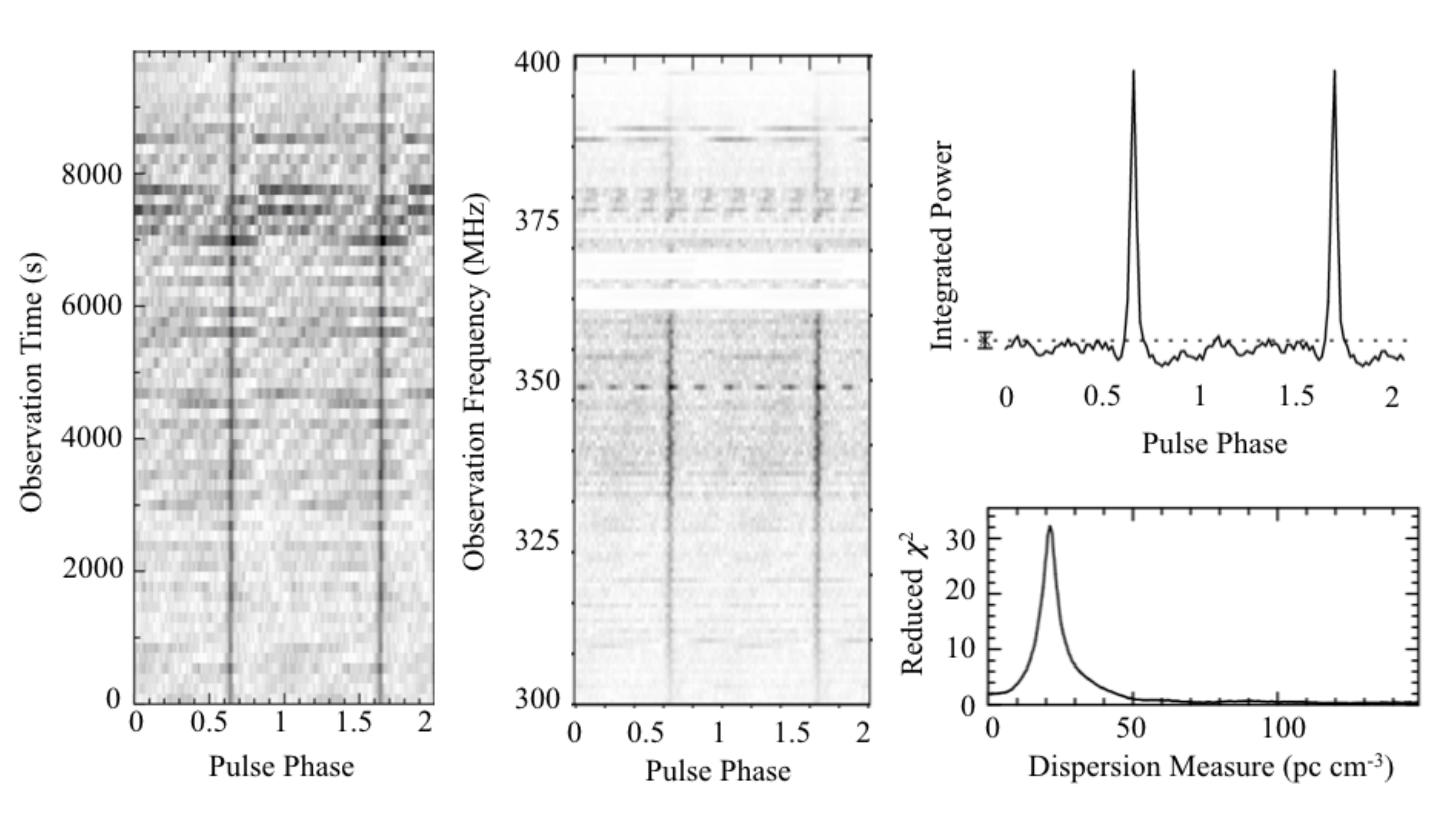}
\caption{ Characteristics of PSR J0802$-$0955. See text for details. \label{fig:1}}
\end{center}
\end{figure}

PSR J0802$-$0955 has a pulse period of 571.2560$\pm$0.0003 ms and a pulse period derivative $<2.1\times$10$^{-10}$ s s$^{-1}$. We provide the radio power folded on the pulse period as a function of both time and radio frequency in Figure \ref{fig:1}, as well as the total integrated pulse profile. Figure \ref{fig:1} additionally shows that the reduced $\chi^2$ metric of the dispersion measure is sharply peaked around 21.3$\pm$0.4 pc cm$^{-3}$ \citep[implying a distance of 1.3 kpc, NE2001 model;][]{cordes02}. Finally, although we are denoting this pulsar by the position at which the GBT was pointing during our observations, the full-width half-max of the GBT beam at 340 MHz is 36\amin, constraining the actual position of this pulsar to a $\approx$1 square degree region around the LMWD SDSS J080250.13$-$095549.8.

\acknowledgments

J.J.A.\ acknowledges funding through the NRAO Student Observing Support program as well as funding from the European Research Council under the European Union's Seventh Framework Programme (FP/2007-2013)/ERC Grant Agreement n. 617001.

\end{document}